\documentclass[aps,prl,twocolumn,groupadress]{revtex4}
\usepackage{graphicx}
\usepackage{amsmath}
\usepackage{amssymb}
\usepackage{epsfig}

\begin{document}

\title{Soft Spin Wave Near $\nu$=1: Evidence for a Magnetic Instability in Skyrmion Systems.}

\author{Yann Gallais$^{1}$}
\email{yann.gallais@univ-paris-diderot.fr}
\altaffiliation{Permanent address: Laboratoire Mat\'eriaux et Ph\'enom\`enes Quantiques, CNRS UMR 7162, Universit\'e Paris 7, France}
\author{Jun Yan$^{1}$}
\author{Aron Pinczuk$^{1,2}$}
\author{Loren N. Pfeiffer$^{2}$}
\author{Ken W. West$^{2}$}

\address{$^1$ Departments of Physics and of Applied Physics and Applied Mathematics, Columbia University, New York, NY 10027, USA \\
    $^2$ Bell Labs, Alcatel-Lucent, Murray Hill, New Jersey 07974, USA}

\begin{abstract}
The ground state of the two dimensional electron gas near $\nu$=1
 is investigated by inelastic light scattering measurements carried down to very low
temperatures. Away from $\nu$=1, the ferromagnetic spin wave
collapses and a new low-energy spin wave emerges below the Zeeman
gap. The emergent spin wave shows soft behavior as its
energy increases with temperature and reaches the Zeeman energy for
temperatures above 2~K. The observed softening indicates an instability of the two dimensional electron gas towards a magnetic order that breaks spin rotational symmetry. We discuss our findings in light
of the possible existence of a Skyrme crystal.
\end{abstract}

\maketitle In condensed matter physics, the competition between
distinct correlated electron ground states follows from a delicate
interplay between dimensionality, strength of Coulomb interactions,
and disorder. Correlated states may possess long range spin and/or
charge order that breaks a symmetry that,  in turn, has profound
consequences on the nature of the low-energy excitation spectrum
\cite{Anderson,Goldstone}. One well-known example of the
relationship between broken-symmetry and elementary excitations is
the isotropic ferromagnet. The broken rotational spin symmetry leads
to appearance of gapless spin wave excitations,  Goldstone modes,
that are fluctuations of the spin density around the magnetization
direction.
\par
The ground state of the two-dimensional (2D) electron gas at Landau
level filling factor $\nu$=1 is a special kind of itinerant
ferromagnet, the quantum Hall ferromagnet, where all electrons
occupy the lowest orbital Landau level and their spins are aligned
along the external magnetic field. The quantum Hall ferromagnet
supports collective excitations similar to ferromagnetic spin-waves
with a gap given by the bare Zeeman energy  E$_z$
\cite{Bychkov,Kallin}. The quantum Hall ferromagnet also has spin
texture excitations built from Skyrmions \cite{Lee,Sondhi}, a concept
used to describe the emergence of nuclear particles in the context
of field theories of nuclear matter \cite{Skyrme}. Skyrmions are
topological objects which smoothly distort the ferromagnetic order
in a vortex-like configuration. Each individual Skyrmion involves
several flipped spins and is therefore not favored by the Zeeman
energy. On the other hand because the 
exchange energy is large in
quantum Hall systems, and it prefers locally aligned spins,
Skyrmions are cheaper than single spin-flips for sufficiently low
E$_z$. The relevance of Skyrmions
has been demonstrated by a wide range of experiments probing the 
spin polarization of the 2DES \cite{Barrett,Schmeller,Maude,Aifer}.
\par
Near $\nu$=1 the interaction between Skyrmions may lock orientations of spin in
the XY plane to favor the formation of a Skyrme crystal of electron
spin orientation that breaks spin rotational symmetry about the
magnetic field axis \cite{Brey}. As a consequence of the additional
symmetry breaking, the 2D electron system supports spin waves which,
contrary to the ferromagnetic spin wave, remain gapless in the
presence of the magnetic field \cite{Sachdev,Cote,Timm}. Both a jump in
the specific heat and the very short T$_1$ observed
away from $\nu$=1 were interpreted as indirect consequences of the
enhanced coupling between the nuclear spins and the electron system due to
the gapless spin waves (Goldstone mode) of the Skyrme crystal
 \cite{Bayot1,Desrat,Gervais}. 
 
 \par
 While most studies of Skyrmions have focused solely on their impact on electron spin polarization, no experiment has explored up to now the possibility of exploring Skyrmion interactions and the magnetic ground state of the 2DES near $\nu$=1 by probing long-wavelength spin wave excitations. Measurements of spin wave excitation are especially valuable since they are expected to reflect the presence of a broken symmetry state arising from XY spin ordering \cite{Cote}.
\par
Here we report the direct observations of spin wave excitations well
below the Zeeman energy. These spin waves emerge near the quantum
Hall state at $\nu$=1. The excitations are measured directly by
inelastic light scattering in experiments that search for
fingerprints of a broken symmetry ground state by looking at the
evolution of spin wave excitation spectra at low temperatures.
\par
Very close to $\nu$=1, the spectrum is dominated by the ferromagnetic
spin wave at energy near  E$_z$. Tuning the filling factor slightly away from
$\nu$=1 at low temperatures T$<$2~K  alters dramatically the spin
excitation spectrum. The ferromagnetic spin wave is strongly
suppressed on both sides of $\nu$=1 signaling the rapid collapse of
the quantum Hall ferromagnetic order. Simultaneously, a new low
energy spin excitation emerges well below the Zeeman energy.
Strikingly the new spin wave displays soft mode behavior with
temperature and a mode at E$_z$ is recovered for T$>$2~K.
\par
Our findings indicate an instability towards a ground state with
spin orientational order at very low temperature that is linked to
Skyrmion interactions. The spin order is consistent with the Skyrme crystal phase
 where the Skyrmions localize on a square lattice and
the XY spins of neighbouring Skyrmions are antiferromagnetically
aligned \cite{Brey}. 

\par
\begin{figure}
\centering \epsfig{figure=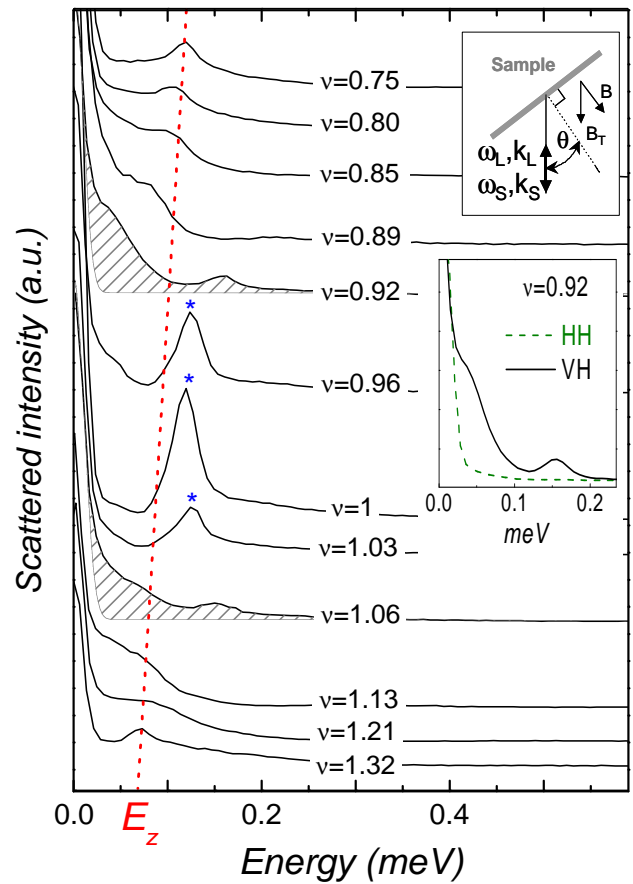, width=0.95\linewidth,clip=}
\caption{(color online) Filling factor dependence of inelastic light
scattering spectra at T$\sim$40~mK. The blue
star marks the ferromagnetic spin wave close to $\nu$=1. The red
dotted line displays the approximate evolution of the bare Zeeman
energy with magnetic field assuming E$_z$=g$\mu_B$B$_T$ with g=-0.44.
The shaded area 
highlights the spectral weight emerging below E$_z$. The insets show
the scattering geometry and the polarization dependence of the
spectrum at $\nu$=0.92. HH and VH indicate configurations in which
the incident and scattered photons are parallel and perpendicular
respectively.} \label{fig1}
\end{figure}

The inelastic light scattering measurements were performed on a high
quality GaAs single quantum well of width 330~$\AA$. Its density is
n=5.5x10$^{10}$~cm$^{-2}$ and its low temperature mobility
$\mu$=7.2x10$^6$~cm$^2$/V s. The magnetic field perpendicular to the
sample is B=B$_T$cos$\theta$ as shown in the top inset of Fig. \ref{fig1}.
The results reported here have been
obtained with $\theta$=50$\pm 2^\circ$. The combination of tilt
angle and electron density gives a ratio of Zeeman to Coulomb
energy, E$_z$/E$_c$=0.0134, where E$_c=e^2/\varepsilon l_o$,
$l_o=(\hbar c/eB)^{1/2}$ is the magnetic length and $\epsilon$ is
the dielectric constant. This ratio lies in the regime where
Skyrmions excitations are cheaper than single spin-flips
\cite{Fertig,Melinte}. The sample was mounted on the cold finger of
a $^3$He/$^4$He dilution refrigerator that is inserted in the cold
bore of a superconducting magnet. The refrigerator is equipped with
windows for optical access. Cold finger temperatures can reach as
low as T=40~mK.
\par
Resonant inelastic light scattering spectra were obtained by tuning
the photon energy of a Ti:Sapphire laser close to the fundamental
optical gap of GaAs. The back scattering geometry was used where the
incident laser beam makes an angle $\theta$ with the normal of the
sample surface (see the top inset of Fig.\ref{fig1}). The wave vector
transferred from the photon to the electron system is
$q=k_L-k_S=(2\omega_L/c)sin\theta$, where $k_{L(S)}$ is the in-plane
component of the wave vector of the incident (scattered) photon.
$q$=1.2x10$^5$~cm$^{-1}$ for $\theta=50^\circ$.
 The power density was kept around 10$^{-5}$~W/cm$^2$ to prevent heating of
the electron gas.
The scattered signal was dispersed by a triple grating
spectrometer working in additive mode and analyzed by a CCD camera
with 26~$\mu m$ wide pixels. At a slit width of 30~$\mu m$, the
combined resolution of the system is about 0.02~meV.
\par
Figure \ref{fig1} displays the filling factor dependence of the
spectrum in the filling factor range 0.75$<\nu<$1.32 at
T$\sim$40~mK. All the features appear predominantly in the
depolarized configuration (VH) which, according to light scattering
selection rules, indicates their spin origin \cite{Yafet}. At $\nu$=1
the spectrum consists of a single sharp (FWHM$\sim$30~$\mu$eV) peak
which is identified as the ferromagnetic spin wave (SW) at the wave vector $q$. Its
energy is blue-shifted from E$_z$ due to the
combined effects of the finite wave vector transfer $q$ and the
large spin-stiffness of the $\nu$=1 quantum Hall ferromagnet
\cite{Gallais}.
\par
Slight departures from $\nu$=1 have major impact on the the
spin wave spectra for \textit{both} $\nu<$1 and $\nu>$1.
The ferromagnetic spin wave intensity is strongly suppressed
leaving a weak and broad peak at higher energies (at $\nu$=0.92 and
$\nu$=1.06 for example). The suppression of the
ferromagnetic spin wave is accompanied by the emergence of a strong
spectral weight \textit{below} E$_z$, with a spectral response that
extends essentially to zero energy. The low-lying spectral weight
appears particularly large for the spectra at $\nu$=0.92 and 1.06 but it is already present,
albeit weaker, for $\nu$=0.96 and $\nu$=1.03. Further away from $\nu$=1, a
well-defined peak at E$_z$ is recovered.

\begin{figure}
\centering \epsfig{figure=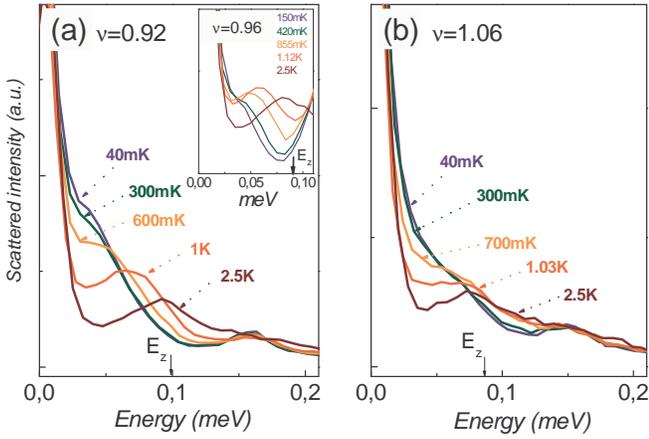, width=0.99\linewidth,clip=}
\caption{(color online) Temperature dependence of the light
scattering spectra at (a) $\nu$=0.92 (B$_T$=4.0~T) and (b) 1.06
(B$_T$=3.5~T). The Zeeman energy, E$_z$, is marked by an arrow for
both filling factors. The inset in (a) shows the temperature dependence of the low
energy spectrum at $\nu$=0.96 (B$_T$=3.85~T).} \label{fig2}
\end{figure}

\par
The collapse of the ferromagnetic spin wave suggests a rapid spin
depolarization away from $\nu$=1 due to the collapse of the spin
order of the ferromagnetic quantum Hall state. The simultaneous
appearance of a low-energy spectral weight on both sides of $\nu$=1
indicates the presence of spin excitations well below the Zeeman
gap in agreement with the very short nuclear spin relaxation times T$_1$ observed
in NMR experiments \cite{Barrett,Tycko,Hashimoto,Desrat,Gervais,Tracy}.
\par
Slightly away from $\nu$=1, the spin excitation spectrum displays
the striking temperature evolution seen in Fig. \ref{fig2} at
$\nu$=0.92 and 1.06 (see also $\nu$=0.96 in the inset). For both filling factors the spin excitation
spectrum at 2.5~K is dominated by a relatively broad peak centered
close to the Zeeman energy. Upon cooling however, the peak
continuously evolves towards lower energy and below approximately
0.6~K most of the spectral weight has already shifted well below
E$_z$ and close the stray light peak centered at zero energy.
\par
\begin{figure}
\centering \epsfig{figure=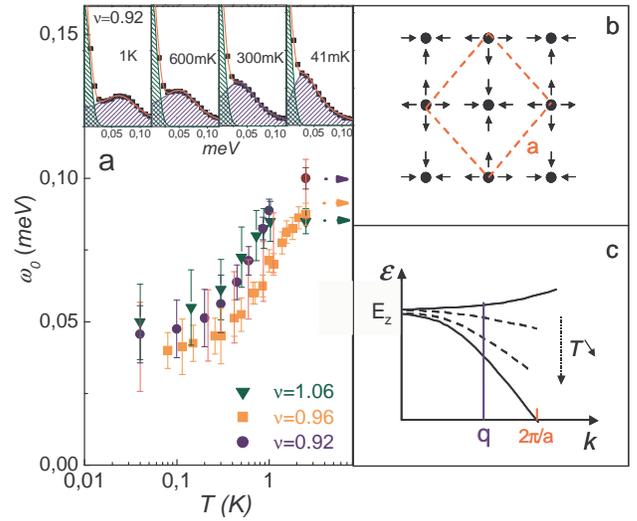, width=0.95\linewidth,clip=}
\caption{(color online) (a) Temperature dependence of the spin wave
energy for $\nu$=1.06, $\nu$=0.96 and $\nu$=0.92. The Zeeman energies for each filling factor are marked by arrows. The inset shows fits of the spectra at $\nu=0.92$ for 4 different
temperatures. The stray light and damped oscillator mode contributions (see text)
are highlighted. (b) A schematic representation of the XY component of the
spin for a square Skyrme crystal with lattice parameter $a$ showing
antiferromagnetic (AF) ordering of the XY spin components of
neighbouring Skyrmions \cite{Brey}. (c) Evolution of the spin wave
dispersion as function of temperature where k$_{sk}$=2$\pi$/a.} \label{fig3}
\end{figure}

Focusing on the low energy part of the spectra
(i.e. $\omega\leq$E$_z$), we have modelled it by a damped oscillator
mode response \cite{Loudon}:
\begin{equation}
\centering
I(T,\omega)\sim[1+n(\omega,T)]\frac{\omega\gamma(T)}{[\omega^2-\omega_0^2(T)]^2+\omega^2\gamma^2(T)}
\end{equation}
where $I(T,\omega)$ is the scattered intensity, $\omega_0(T)$ the
mode energy, $\gamma(T)$ the damping and n($\omega,T)$ is the
Bose-Einstein factor. The stray light central peak was fitted using
a gaussian profile centered at $\omega$=0 (see Fig.\ref{fig3}). 
The spin wave remains overdamped at all temperatures and its energy continuously evolves towards lower energies upon cooling as shown in Fig. \ref{fig3}(a).
\par
The striking softening of the spin wave that occurs at temperatures
below 2.5~K seen in Figs. \ref{fig2} and \ref{fig3}(a) shows a
trend towards a magnetic instability for filling factors in the
range $|\nu-1|\sim[0.04;0.10]$. From a gapped spectrum consistent with ferromagnetic spin order along the $z$ direction (magnetic field direction) only, the system gradually evolves upon cooling  towards a spectrum with essentially gapless spin excitations indicating a state with additional in-plane XY spin order \cite{Sachdev}. In this scenario the softening of the spin wave below the Zeeman energy is connected to a \textit{spontaneous breaking of spin rotational symmetry} that is
consistent with the antiferromagnetic (AF) alignment of XY spin
components like the one predicted for a Skyrme crystal phase and shown in Fig. \ref{fig3}(b)
\cite{Brey,Cote}.
\par
The softening of the spin wave is expected to occur around the reciprocal 
lattice wave-vector $k_{sk}$ of the new XY AF magnetic lattice sketched in Fig. \ref{fig3}(c).
A proposed temperature evolution of the spin wave dispersion is depicted in Fig. \ref{fig3}(d): starting from a conventionnal ferromagnetic quadratic dispersion which is gapped at E$_z$ the dispersion shows a developping anomaly at the reciprocal magnetic lattice wave-vector, $k_{sk}$ as
the temperature is decreased.
We emphasize here that the sensitivity of our experiment to the magnetic
instability is crucially linked to the fact that we are probing the
 spin wave excitation spectrum at a
wave vector $q$ that is in the range of the expected lattice
wave vector of the square Skyrme crystal in the filling factor
region of interest \footnote{A straightforward calculation gives
$q\sim0.4k_{sk}$ for $\nu$=0.92 in our sample.}.
\par
While our data are strong evidences for a ground state with broken spin rotational symmetry that is consistent with a Skyrme crystal picture,
we cannot rule out however a correlated Skyrme liquid phase with long range
spin orientationnal order only and no long range translational order.
Both states, liquid ans crystal, are expected to have a similar spin excitation spectrum \cite{Cote}.

\par
\begin{figure}
\centering \epsfig{figure=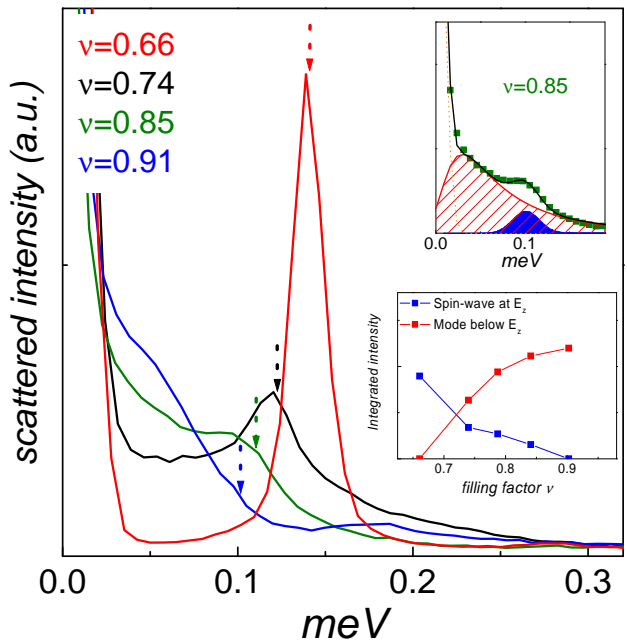, width=0.95\linewidth,clip=}
\caption{(color online) Filling factor dependence of the low energy
spectrum at T=40~mK. The bare Zeeman energy is marked by an arrow
for each filling factors. For $\nu<0.85$ the additional spin wave at
E$_z$ was modelled by a gaussian profile. The inset shows an example
of fit for $\nu$=0.85 with the spin wave below E$_z$ in red and the
ferromagnetic spin wave contribution in blue. Also shown as inset is
the filling factor dependence of the spectral weights of the two
contributions.} \label{fig4}
\end{figure}

Figure \ref{fig4} shows the evolution of the spin excitation spectrum
at 40~mK for $\nu<0.9$. When moving further away from $\nu$=1, the new low
energy the spin wave broadens significantly and the spectral
weight below E$_z$ decreases drastically. Simultaneously a
well-defined spin wave, centered at E$_z$, develops
and the spin excitation spectra become almost temperature
independent below 1~K for $\nu<0.85$ (not shown). When $\nu$=2/3
is reached the spectrum is gapped with vanishing spectral weight below
E$_z$. It displays a well-defined ferromagnetic spin wave similar to $\nu$=1,
suggesting a spin polarized quantum
Hall fluid at this filling factor.
\par
 The filling factor dependence of the spectral weight of both the spin wave
below E$_z$ and the ferromagnetic spin wave at E$_z$ shown in the bottom inset of Fig. \ref{fig4} suggests a crossover between spin polarized quantum Hall fluids
with gapped spin-flip excitations and Skyrmions ground states in the
filling factor range $2/3<\nu<1$ \cite{Nayak,Paredes}. The behavior
can be qualitatively understood as follows: as the density of
Skyrmions increases upon decreasing filling factor, their size
decreases until they become equivalent to single spin-flips and
conventional fractionnal quantum Hall fluids emerge.
\par

To conclude, we have reported direct measurement of the low energy spin
excitation spectrum around $\nu$=1. Slightly away from $\nu$=1, a
low-lying soft spin wave is observed, indicating a ground state with broken spin rotational symmetry that arises from magnetic interaction between neighbouring Skyrmions. 
\par
This work is supported by the NSF (Grant No. NMR-0352738) and DOE
(Grant No.
DE-AIO2-04ER46133), and by a research grant from the W.M. Keck
Foundation. Y.G. acknowledges partial support from a CNRS-USA travel grant.
The authors would like to thank L. Brey, N. R. Cooper, R. C\^ot\'e and C. Tejedor for fruitfull discussions and B. S. Dennis for valuable advice.

\end{document}